\documentclass[12pt,prl,nofootinbib,superscriptaddress]{revtex4-1}

\usepackage[utf8]{inputenc}
\usepackage{amssymb}
\usepackage{amsmath}
\usepackage{graphicx}
\usepackage{psfrag}
\usepackage{latexsym}

\newcommand{\p}{\partial}

\newcommand{\bea}{\begin{eqnarray}}
\newcommand{\beal}[1]{\begin{eqnarray}\label{#1}}
\newcommand{\eea}{\end{eqnarray}} 
\newcommand{\be}{\begin{equation}} 
\newcommand{\bel}[1]{\begin{equation}\label{#1}}
\newcommand{\ee}{\end{equation}} 
\newcommand{\rf}[1]{(\ref{#1})}
\newcommand{\f}[2]{\frac{#1}{#2}}
\newcommand{\nn}{\nonumber}
\newcommand{\bit}{\begin{itemize}}
\newcommand{\eit}{\end{itemize}}
\newcommand{\ben}{\begin{enumerate}}
\newcommand{\een}{\end{enumerate}}

\newcommand{\symm}{${\mathcal N}=4$ SYM}

\begin{document}

\title{Hydrodynamics beyond the Gradient Expansion: Resurgence and Resummation}

\author{Michal P. Heller}
\email{mheller@perimeterinstitute.ca}
\affiliation{\it Perimeter Institute for Theoretical Physics, Waterloo, Ontario N2L 2Y5, Canada} 
\affiliation{National Centre for Nuclear Research,  Ho{\.z}a 69, 00-681 Warsaw, Poland}

\author{Micha\l\ Spali\'nski}
% \email{mspal@fuw.edu.pl}
\email{michal.spalinski@ncbj.gov.pl}
\affiliation{National Centre for Nuclear Research,  Ho{\.z}a 69, 00-681 Warsaw, Poland}
\affiliation{Physics Department, University of Bia{\l}ystok, Lipowa 41, 15-424 Bia{\l}ystok, Poland}

\begin{abstract}

Consistent formulations of relativistic viscous hydrodynamics involve
short-lived modes, leading to asymptotic rather than convergent gradient
expansions. 
In this Letter we consider the
M{\"u}ller-Israel-Stewart theory applied to a longitudinally expanding
quark-gluon plasma system and identify hydrodynamics as a universal attractor
without invoking the gradient expansion. We give strong evidence for the
existence of this attractor and then show that it can be recovered from the
divergent gradient expansion by Borel summation. This requires careful
accounting for the short-lived modes which leads to an intricate mathematical
structure known from the theory of resurgence.

\end{abstract}

\maketitle

\noindent \emph{Introduction.--} The past 15 years have witnessed the rising
practical importance of relativistic viscous hydrodynamics. One reason for 
this is the success of hydrodynamic modeling of quark-gluon plasma (QGP) in
heavy ion collision experiments at the RHIC and the LHC and the realization that QGP
viscosity provides a crucial probe of QCD physics
\cite{Shuryak:2014zxa}. Another motivation is the relation between black holes
and fluids, which originated in the 1970s as the rather mysterious black hole
membrane paradigm~\cite{Damour:1978cg}. With the advent of holography, this
connection has been promoted to a precise
correspondence~\cite{Bhattacharyya:2008jc} shedding light both on the physics
of QGP and 
liquids in general, as well as gravity.

The perfect fluid approximation is widely used in astrophysics and this
theoretical description of relativistic inviscid fluids is rather
well established~\cite{Andersson:2006nr}. On the the other hand, relativistic
viscous hydrodynamics is much less well understood. One of the recent insights
is to regard hydrodynamics as a systematic gradient expansion, much in the
spirit of low-energy effective field theory~\cite{Baier:2007ix}. 

However, the
requirement of causality leads to a framework which necessarily incorporates
very large momenta (and frequencies). In all known examples this is
accompanied by the appearance of short-lived excitations -- nonhydrodynamic
modes. It has recently been 
shown, in the context of AdS/CFT correspondence, that their presence leads to
the divergence of the hydrodynamic gradient series for strongly coupled ${\cal
  N} = 4$ super Yang-Mills (SYM) plasma~\cite{Heller:2013fn}. In view of this
it is not 
clear whether or how a naive gradient expansion defines the theory. This is in
fact a fundamental conceptual question concerning relativistic hydrodynamics
as such.

In this Letter we propose a definite answer: since the nonhydrodynamic modes
decay exponentially, the system relaxes to an attractor regardless of when an
initial condition is set. In the following we consider a simple situation in
which this can be made completely explicit: the M{\"u}ller-Israel-Stewart
(MIS) theory \cite{Muller:1967zza,Israel:1979wp,Baier:2007ix} specialized to a
longitudinally expanding conformal fluid. We show that the attractor can be
determined by relaxation from solutions which take the form of a
{\em transseries}. The higher orders of this transseries are encoded in the
divergent hydrodynamic gradient expansion, in line with expectations 
based on resurgence ideas~\cite{Dunne}.

\vspace{10 pt}

\noindent \emph{M{\"u}ller-Israel-Stewart theory.--} The Landau-Lifschitz
formulation of relativistic viscous hydrodynamics~\cite{LLfluid} asserts that
the evolution equations for the hydrodynamic fields -- temperature $T$ and
flow velocity $u^{\mu}$ -- are the conservation equations of the
energy-momentum tensor 
\bel{hydro}
\langle T^{\mu \nu} \rangle = {\cal E} \,  u^{\mu} u^{\nu} +  {\cal P} ({\cal E}) (
\eta^{\mu \nu} + u^{\mu} u^{\nu} ) + \Pi^{\mu \nu},
\ee
where the shear stress tensor $\Pi^{\mu \nu}$ is given by
% no bulk viscosity (maybe comment on this)
\bel{PiLL}
\Pi^{\mu \nu} = - \eta \sigma^{\mu \nu}.
\ee
From a modern perspective, one could contemplate including on the RHS of
Eq.~\rf{PiLL} all 
possible terms graded by the number of derivatives of $T$ and $u^{\mu}$. If
this is done to a finite order, as in Eq.~\rf{PiLL}, the resulting theory will
not have a well-posed initial value problem due to superluminal signal
propagation~\cite{Hiscock:1985zz,PhysRevD.62.023003,Geroch:1995bx,Geroch:2001xs}.

MIS theory resolves this problem by promoting the shear stress tensor to an 
independent dynamical 
field which satisfies a relaxation-type equation: 
\bel{eqMIS}
\left( \tau_\Pi u^\alpha\p_\alpha  + 1 \right) \Pi^{\mu \nu} = - \eta
\sigma^{\mu\nu} + \ldots  \, ,
\ee
where $\tau_{\Pi}$ is a phenomenological parameter (the relaxation time) and
the ellipsis denotes several additional terms whose 
explicit form can be found in~\cite{Baier:2007ix}. 
Linearization of the resulting theory is causal as 
long as $T \tau_{\Pi}\geq \eta/s$. 
This approach has enjoyed great success in describing the
evolution of QGP~\cite{Luzum:2008cw}. 
It has also been obtained as the
long-wavelength effective description of strongly coupled ${\cal N} = 4$ SYM
plasma in the framework of the AdS/CFT
correspondence~\cite{Baier:2007ix,Bhattacharyya:2008jc}. 

The iterative solution of Eq.~\rf{eqMIS} generates the gradient expansion of the
shear stress tensor leading to the appearance of an infinite number of terms
on the RHS of Eq.~\rf{PiLL}. 
Their coefficients (the transport
coefficients of all orders) are expressed in terms of $\eta$ and  
$\tau_{\Pi}$ (and 4 more parameters in the general conformal
case~\cite{Baier:2007ix}).

In the language of high energy physics, MIS theory can be regarded as an UV
completion of Landau-Lifschitz theory. This is in contrast to the standard way
of viewing it as a phenomenological model providing an effective description
of some microscopic system only at late times. By treating MIS theory as an UV
completion we mean that we consider a hypothetical physical system such that
the MIS theory describes it {\em also at early times}. This could be the case
for systems where the approach to equilibrium is governed by a purely damped
quasinormal mode.

\vspace{10 pt}

\noindent \emph{The setup.--} To overcome the complexity of the MIS equations,
we focus on the case of Bjorken flow~\cite{Bjorken:1982qr}, 
which, due to a very high degree of symmetry, reduces them to a set
of ordinary differential equations. 
The symmetry in question, boost
invariance, can be taken to mean that in proper time-rapidity coordinates
$\tau, y$ (related to Minkowski coordinates $t, z$ by $t = \tau \cosh y$ and
$z = \tau \sinh y$), the energy density, flow velocity and 
shear stress tensor depend only on the proper time $\tau$. The
MIS equations then take the simple form 
\beal{miseqn}
\tau  \dot{\epsilon} &=& - \frac{4}{3}\epsilon + \phi\nonumber\, , \\
\tau_\Pi \dot{\phi} &=& 
\frac{4 \eta}{3 \tau } 
- \frac{\lambda_1\phi^2}{2 \eta^2}
- \frac{4 \tau_\Pi\phi}{3 \tau }
- \phi \, ,
\eea
where the dot denotes a proper time derivative 
and $\phi\equiv-\Pi^{y}_{y}$,
the single independent component of the shear stress 
tensor. 
The term involving $\lambda_1$ comes from the elided terms in
Eq.~\rf{eqMIS}; for details see~\cite{Baier:2007ix}. 

In a conformal theory, $\epsilon \sim T^4$ and 
the transport coefficients satisfy
\be
\tau_\Pi = \frac{ C_{\tau \Pi }}{T}, \qquad \lambda_1 =  C_{\lambda_1}
\frac{\eta}{T}, \qquad \eta = C_\eta\ s \, ,
\ee
where $s$ is the entropy density and $C_{\tau \Pi }, C_{\lambda_1}, C_\eta$ are
dimensionless constants. In the case of  
\symm\ theory their values are known from fluid-gravity duality~\cite{Bhattacharyya:2008jc}:
\bel{symvalues}
 C_{\tau \Pi } = \frac{2-\log (2)}{2 \pi} , \qquad  C_{\lambda_1} = \frac{1}{2
     \pi}, \qquad  C_\eta = \frac{1}{4 \pi}.
\ee

\vspace{10 pt}

\noindent \emph{The hydrodynamic attractor.--} 
From Eq.~\rf{miseqn} one can derive a single second order equation for the
energy density or, equivalently, the temperature:
\beal{misevol}
\tau  C_{\tau \Pi } \frac{\ddot{T}}{T} &+& 3 \tau C_{\tau \Pi }\left(\frac{\dot{T}}{T}\right)^2 
+ (\frac{11 C_{\tau \Pi}}{3 T}+\tau) \dot{T} + \nonumber\\
&-&\frac{4 C_{\eta }}{9 \tau }
+\frac{4 C_{\tau \Pi }}{9 \tau   }
+\frac{1}{3}T = 0.
\eea
To simplify the presentation, we have set
\mbox{$C_{\lambda_1}=0$} in this equation as well as in Eqs.~\rf{feqn},
\rf{fhydro}, and \rf{slowroll} below.

To proceed further it is crucial to rewrite 
Eq.~\rf{misevol} in first order form. Introducing the dimensionless variables
$w$ and $f$ (as in Ref.~\cite{Heller:2011ju}),
\bel{fdef}
w = \tau T, \qquad f = \tau \frac{\dot{w}}{w} \, ,
\ee
the MIS evolution equation \rf{misevol} takes the form 
\beal{feqn}
C_{\tau \Pi } w f f' &+& 4 C_{\tau \Pi} f^2 + 
\left(w -\frac{16 C_{\tau \Pi}}{3}\right) f + \nonumber\\ 
&-&\frac{4 C_{\eta }}{9}+\frac{16 C_{\tau \Pi }}{9}-\frac{2 w}{3} = 0,
\eea
where the prime denotes a derivative with respect to $w$. Equations
\rf{fdef} and \rf{feqn} together are equivalent to Eq.~\rf{misevol} as 
long as the function $w(\tau)$ is invertible.

At large times (which translate to large $w$) we expect universal hydrodynamic
behaviour~\cite{Heller:2011ju}. 
In phenomenological analysis of heavy ion
experiments, usually based on MIS theory, hydrodynamic codes are initialized
typically at $w\approx 0.5$, which corresponds roughly to
a time $\tau = 0.5$ fm after the collision, with 
the temperature $T =$ 350 MeV at the centre of the fireball at RHIC (see
e.g. \cite{Broniowski:2008vp}). Equation \rf{feqn} indeed 
possesses a unique stable 
solution which can be presented as a series in powers of $1/w$:
\bel{fhydro}
f(w) = \frac{2}{3}+\frac{4 C_\eta}{9 w}+\frac{8 C_\eta C_{\tau \Pi}}{27
   w^2}+ O(\frac{1}{w^3}).
\ee
This is, in fact, the hydrodynamic gradient expansion.

It is easy to see that linear perturbations around this formal solution decay
exponentially on a time scale set by $\tau_\Pi$:
\bel{linpert}
\delta f(w) \sim \exp\left(-\frac{3}{2 C_{\tau \Pi}} w\right)
w^{\frac{C_\eta- 2  C_{\lambda_1} }{C_{\tau \Pi}}} 
\left(1+ O(\frac{1}{w})\right) 
\ee
% (in this formula we have included the contribution from $C_{\lambda_1}$). 
This is precisely the short-lived mode introduced by the MIS prescription. In
the language of the gravity dual to \symm\ theory this would be an
analog of a quasinormal mode~\cite{Heller:2013fn,Heller:2014wfa} whose
frequency is purely imaginary.  

The presence of this exponentially decaying mode suggests that
%, at least for large values of $w$, 
Eq.~\rf{feqn} possesses an attractor solution. We propose
that this attractor constitutes the 
definition of hydrodynamic behaviour. 
As discussed below, the presence of this attractor can be inferred without
reference to the gradient expansion, which, as shown in the following section,
is in fact divergent. 

The existence of the hydrodynamic attractor is supported by examining the
behaviour of generic solutions of 
Eq.~\rf{fdef}, with initial conditions set at various values of~$w$. 
\begin{figure}[ht]
\includegraphics[height=0.3\textheight]{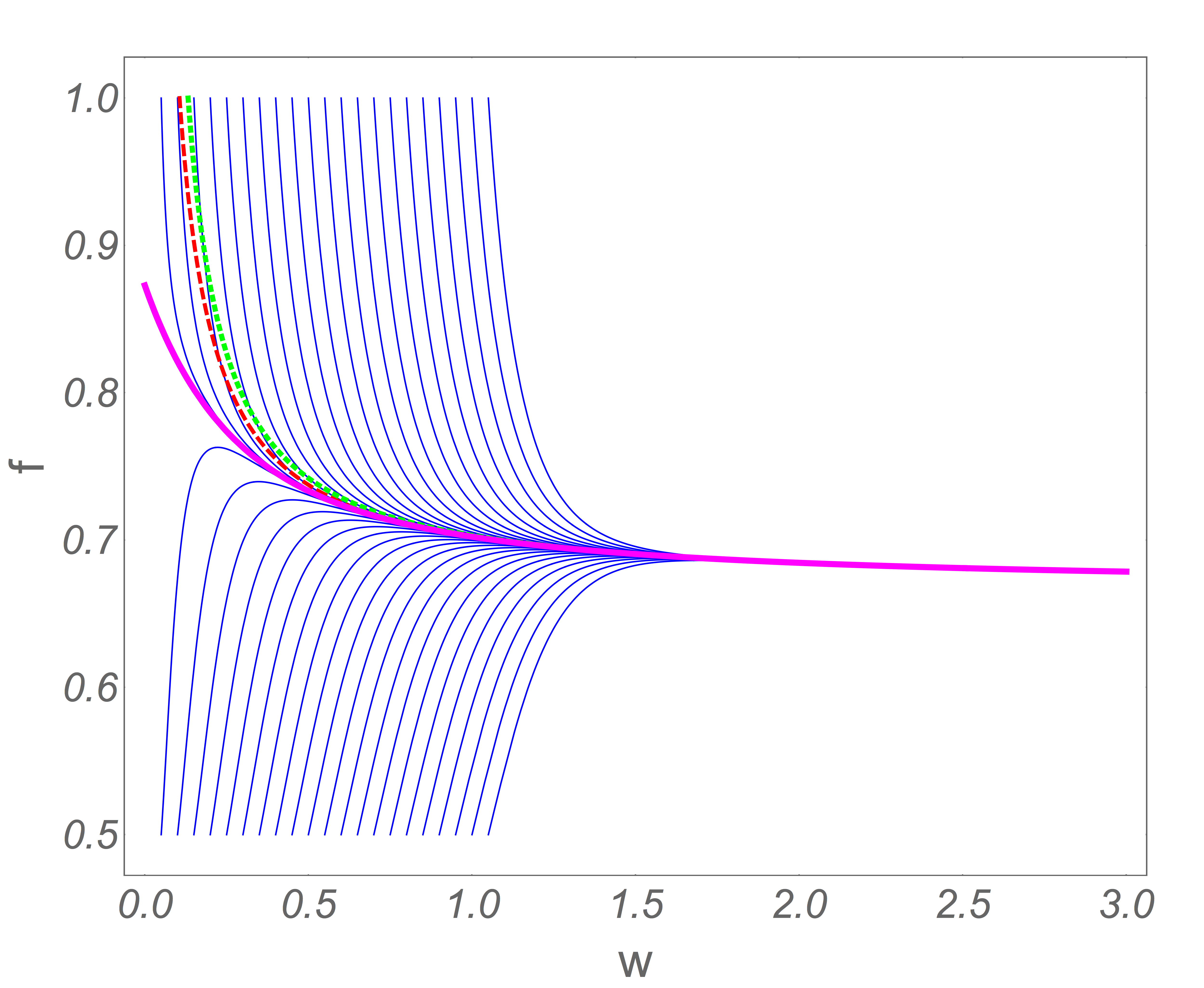}
\includegraphics[height=0.3\textheight]{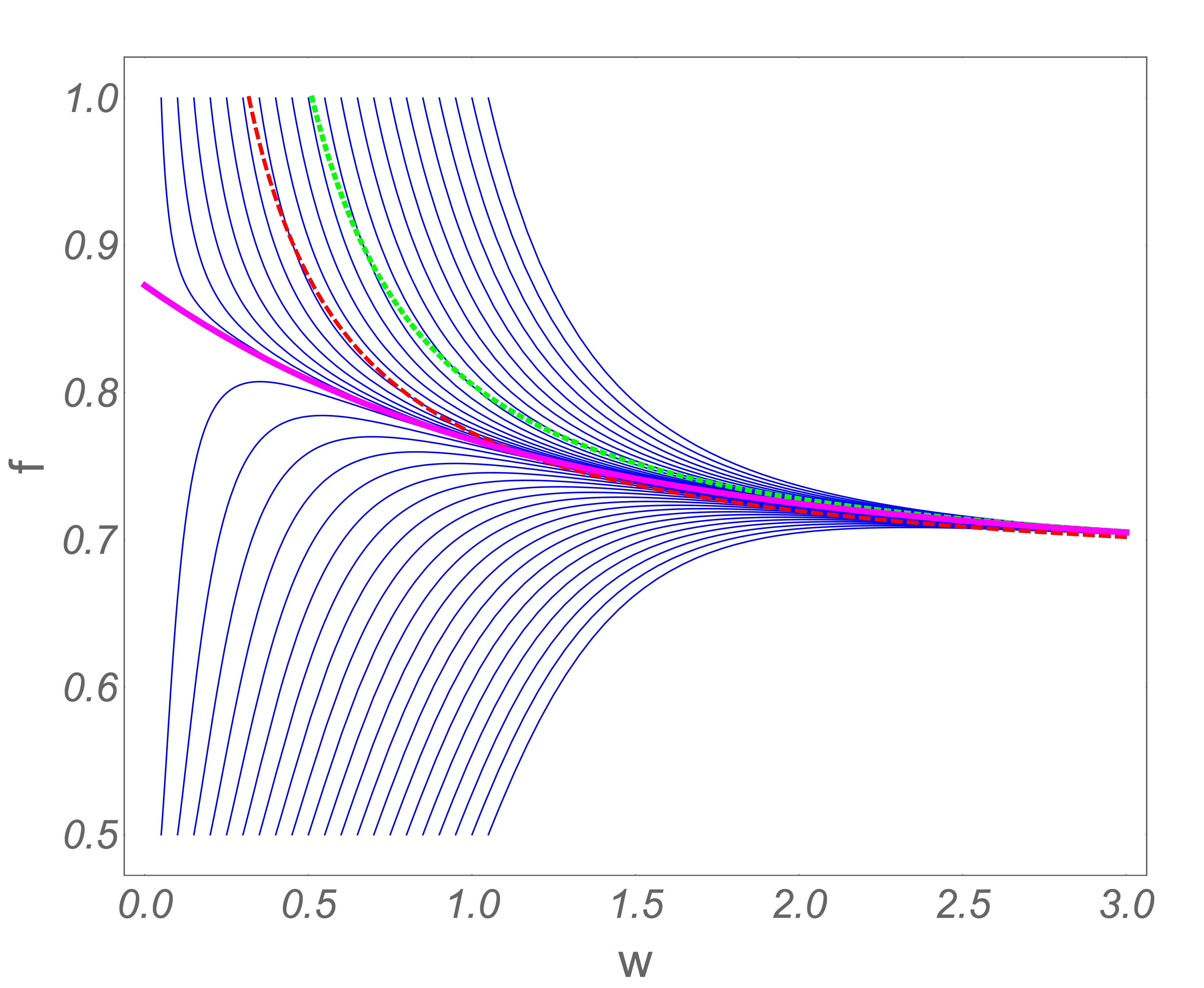}
\caption{The blue lines are numerical solutions of Eq.~\rf{fdef} 
the for various initial conditions; the thick, magenta line is the numerically determined
  attractor. The red, dashed and green, dotted lines represents first and
  second order order hydrodynamics. 
% The plot on the left was made with parameter
The upper plot was made with parameter 
values appropriate for \symm\ theory, 
% while the plot on the right 
while the lower plot 
has both $\eta/s$ and $C_{\tau \Pi }$ increased by a factor of
  3. Note that in the latter case the hydrodynamic attractor is 
  attained at larger values of $w$, as expected.
} 
\label{numa}
\end{figure} 

As seen in Fig.~\ref{numa}, a generic solution rapidly decays to the
attractor. Furthermore, the attractor appears to persist even at very small
values of~$w$, where hydrodynamics of finite order becomes
ill defined. Perhaps unsurprisingly, truncating Eq.~\rf{fhydro} at first or
second order gives results distinctly different from the attractor at very
small~$w$. The magnitude of this difference depends on the 
values of the transport coefficients. 

Assuming \symm\ parameter values, we see
that adopting just the viscous 
hydrodynamics constitutive relations provides a remarkably good
approximation of the attractor for a
wide range of $w$. In particular, this
holds with an error
smaller than $10 \%$ for $w > 0.5$. 

Examining the behaviour of $f$ for $w\approx 0$ one finds two
solutions, one of which is stable
\bel{smallw}
f(w) = \frac{2 \sqrt{C_{\tau \Pi }}+\sqrt{C_{\eta }}}{3
   \sqrt{C_{\tau \Pi }}} + O(w) .
\ee
By setting the initial value of $f$ at $w\approx 0$ arbitrarily close to the
value given by Eq.~\rf{smallw}, the attractor can be 
determined numerically with the result shown in Fig.~\ref{numa}. 

Another way of characterizing the attractor is to expand Eq.~\rf{feqn} in
derivatives of $f$ -- this is an analog of the slow-roll expansion in theories
of inflation (see e.g.~\cite{Liddle:1994dx}). 
This leads to a kind of gradient expansion, but this is not the usual
hydrodynamic expansion, since the generated approximations to $f$ are not 
polynomials in $1/w$. By choosing the correct branch of the square
root which appears at leading order one can ensure that expanding the $k$-th
approximation in powers of $1/w$
one finds consistency with hydro at order $k+1$. 
At leading order one finds
\bel{slowroll}
f(w) = \frac{2}{3} -\frac{w}{8 C_{\tau \Pi}}+ \frac{\sqrt{64 C_\eta C_{\tau
      \Pi} + 9 w^2}}{24 C_{\tau\Pi}} \, .
\ee
Continuing this to second order gives an analytic representation of the
attractor which matches the numerically computed curve 
even for $w$ as small as $0.1$. 
%% final
%% The slow-roll expansion at low orders gives a very
%% accurate representation of the hydro attractor, but it is easily checked that
%% this expansion is also divergent.
%% endfinal

Finally, one can also construct the attractor in an expansion around $w = 0$
starting with the $f(w)$ given by Eq.~\rf{smallw}. It turns out that the radius of
convergence of this series is finite. All three expansion schemes 
are consistent with the numerically determined attractor.

\vspace{10 pt}

\begin{figure}[ht]
\includegraphics[height=0.32\textheight]{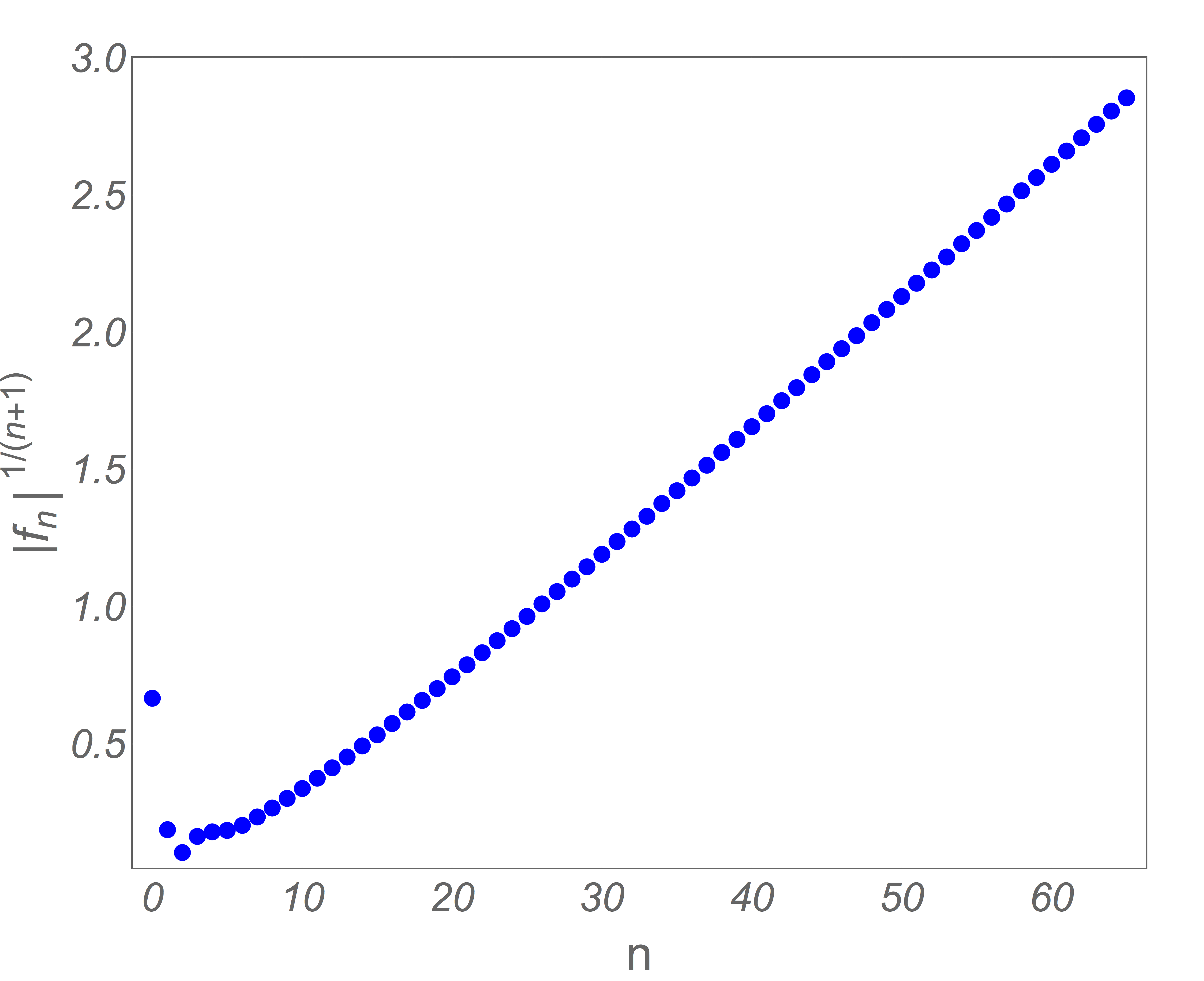}
\caption{The large order behaviour of the hydrodynamic series. The slope
is consistent with location of the singularity nearest to the origin as
given by Eq.~\rf{ratio}. 
%Depending on the value of $C_{\lambda_1}$ the sign of
%the coefficients may change at some value of $n$, but the series is not
%alternating at high order (for \symm\ parameters this happens at $n=3$).
} 
\label{diverge}
\end{figure}

\noindent \emph{Hydrodynamic gradient expansion at high orders.--} In what
follows we focus on the 
hydrodynamic expansion, the expansion in powers of $1/w$. 
It is straightforward to generate the gradient expansion up to essentially
arbitrarily high order (in practice, we chose to stop at 200). 
The coefficients $f_n$ of the series solution 
\bel{eq.fhydro}
f(w)=\sum_{n=0}^\infty f_n w^{-n}
\ee
show factorial behaviour at large $n$, as seen in Fig.~\ref{diverge}.
This is analogous to the
results obtained in Ref.~\cite{Heller:2013fn} for the case of \symm\ theory. 

In view of the divergence of the hydrodynamic expansion, we turn to the Borel
summation technique. The Borel transform 
of $f$ is given by
\bel{borel}
f_B(\xi) = \sum_{n=0}^\infty \frac{f_n}{n!} \xi^n
\ee
and results in a series which has a finite radius of convergence. Note that in
Eq. \rf{borel} large $w$ corresponds to small $\xi$. To invert the Borel
transform, it is necessary to know the analytic continuation of the series
\rf{borel}, which we denote by $\tilde{f}_{B}(\xi)$. The inverse Borel 
transform 
\be
\label{eq.borelsum}
f_{R}(w) = \int_{C} d\xi \, e^{-\xi} \, \tilde{f}_B(\xi/w) =  w \int_{C} d\xi
\, e^{- w \xi} \, \tilde{f}_B(\xi)  
\ee
where $C$ denotes a contour in the complex plane connecting $0$ and $\infty$,
is interpreted as a resummation of the original divergent series
\rf{eq.fhydro}. To carry out the integration, it is essential to know the
analytic structure of $\tilde{f}_B(\xi)$.

We perform the analytic continuation using diagonal Pad\'e
approximants~\cite{PhysRevB.7.3346}, given by the 
ratio of two polynomials of order 100. This function has a dense sequence of
poles on the real axis, starting at $\xi_0=7.21187$, which signals the
presence of a cut originating at that point~\cite{PadeCut}. This can be
corroborated by applying the ratio method~\cite{PhysRevB.7.3346}, which allows
the estimation of the location and order of the leading branch-cut singularity
by examining the series coefficients. Specifically, if the function
approximated by Eq.~\rf{borel} has the leading singularity of the form 
$(\xi_0 -\xi)^\gamma$ then for large $n$
\bel{ratio}
% \frac{f_n}{f_{n+1}} = \xi_0 \frac{n}{n+1} \left(1 + \frac{1+\gamma}{n} +
% O(\frac{1}{n^2})\right).
\frac{f_n}{f_{n+1}} = \frac{ \xi_0}{n+1} \left(1 + \frac{\gamma+1}{n+1} + O(\frac{1}{n^2})\right).
\ee
Applying this formula one finds $\xi_0=7.21181$ 
%%%PRL, which is consistent with the result obtained using the Pad\'e
%%%approximant, and $\gamma=1.1449$.  
(which is consistent with the pattern of poles) and  $\gamma=1.1449$. 

%%%PRL In fact, we 
As we argue in the Supplemental Material, the analytic structure of
%%%PRL the analytic continuation 
$\tilde{f}_B$ must involve further singularities on the real axis precisely in
the following form: 
\bel{eq.g}
\tilde{f}_B(\xi) = h_{0}(\xi) + (\xi_{0} - \xi)^{\gamma}
h_1(\xi) +  (2 \xi_{0} - \xi)^{2\gamma} h_2(\xi) + 
\dots 
\ee
where the functions $h_{k}(\xi)$ are analytic and the ellipsis denotes
further singularities at integer multiples of $\xi_0$.

Such branch-cut singularities of $\tilde{f}_{B}$ lead to ambiguities in the
inverse Borel transform. Indeed, such a series is not Borel summable in the
usual sense. It is, however, known that even in such cases a resummation is
possible (see, e.g., Ref.~\cite{2012arXiv1210.3554A}), but requires a
nontrivial choice of integration contour. The freedom in the choice of
integration 
contour leads to complex ambiguities
\be
\label{eq.ambiguity}
\delta f_R(w) = e^{i \pi k p \gamma} w \int_{k \xi_{0}}^{\infty} d\xi \,
e^{-w \, \xi}(\xi - k\xi_{0})^{k \gamma} h_k(\xi) , 
\ee
where $p$ is an odd integer reflecting the choice of Riemann sheet. For large
$w$ this becomes 
\bel{ambiguity}
\delta f_R(w)  \approx e^{i \pi k p \gamma} \Gamma(k \gamma+1)
h_k(k\xi_{0}) \left(w^{- \gamma} e^{- w \xi_{0}}\right)^k   \, .
\ee
This ambiguity is a feature of the hydrodynamic series and its presence is an
indication of physics outside the gradient expansion. We saw in the previous
section that there are nonanalytic, exponentially suppressed corrections to
the hydrodynamic series following from the presence of the nonhydrodynamic
MIS mode. These have precisely the correct structure to eliminate the $k=1$ 
ambiguity in inverting the Borel transform. Indeed, comparing
Eq. \rf{ambiguity} with Eq.~\rf{linpert} we are led to identify $\xi_0$ with
$3/2C_{\tau\Pi}$ and $-\gamma$ with $(C_\eta-2  C_{\lambda_1})/C_{\tau\Pi}$. Evaluating these
combinations with parameter values appropriate for \symm\ theory
[Eq.~\rf{symvalues}] gives agreement to five significant digits. Both 
Eq.~\rf{ambiguity} and 
Eq.~\rf{linpert} receive corrections in $1/w$ and we expect them to match
also.

The nonlinear structure of Eq.~\rf{feqn} suggests the presence of an infinite
series of exponential corrections, which are matched by further branch
cuts in Eq.~\rf{eq.g}. In the following
section we calculate these corrections and give strong evidence that they
conspire to yield an unambiguous, finite and real answer for $f_R$, up to a
real constant of integration.

\vspace{10 pt}

\noindent \emph{Resurgence --} The results presented so far
suggest that Eq.~\rf{feqn} should possess a solution in the form of a
transseries \cite{Dorigoni:2014hea}: 
\bel{eq.transs}
f(w) = \sum_{m = 0}^{\infty} c^m  \Omega(w)^m \sum_{n = 0}^{\infty} a_{m, n}
w^{-n} \, ,
\ee
where $\Omega\equiv w^{-\gamma}\exp(-w\xi_0)$ while $c$ and $a_{m,n}$ are
coefficients to be determined by the equation.  
%%PRL Indeed, by direct substitution one can check that all the coefficients
%%$a_{m,n}$ in 
%% Eq.~\rf{eq.transs} are fixed uniquely apart from  $a_{1,0}$, which can be
%% absorbed into the constant $c$.
By direct substitution one can check up to high order that all of the
coefficients $a_{m,n}$ in Eq.~\rf{eq.transs} are fixed uniquely apart from
$a_{1,0}$, which can be  absorbed into the constant $c$.

For each value of $m$ in Eq.~\rf{eq.transs}, the series over $n$ is expected to
be divergent -- we have checked this for $m\leq 2$. Applying the Pad\'e-Borel
techniques discussed 
% in the previous section 
earlier leads to complex resummation ambiguities for each 
of these series. To obtain a meaningful answer, it must be 
possible to choose the single complex constant $c$ in such a way that the
result does not depend on the choice of integration contours and that the
imaginary parts cancel.

The key observation is that the ambiguity at the leading order of the
transseries is 
proportional to $\Omega$, so it can only be canceled by terms of order
$m=1$ or higher. This cancellation determines the constant~$c$ 
%%PRL The condition for canceling the ambiguity at first order in $\Omega$
%% determines the constant~$c$ 
\bel{csol}
c = r - e^{i \pi \gamma p} \Gamma (\gamma +1) h^{(0)}_1(\xi_{0})
\ee
up to an arbitrary real number $r$, which is the expected integration constant
for the first order differential equation \rf{feqn}. 
%%PRL Alternatively, one may say that the cancellation of ambiguities
%%determines the imaginary part of~$c$. 

\vspace{10 pt}

\noindent \emph{Resummation --} Having provided strong evidence for the
existence of an unambiguous and physically sensible result encoded in the
transseries, we now invert the Borel sums for $m\leq 2$. For these
calculations we used extended precision arithmetic (keeping one thousand
digits). 
%%%PRL One should be aware of various limitations of this calculation which
%%%arise from transseries truncation errors and systematic errors introduced
%%%by the analytic continuation (note that we used extended precision
%%%arithmetic keeping one thousand digits).  

Inverting the Borel transform at each order of the transseries requires
performing the integration in Eq.~\rf{eq.borelsum}.  The analytic continuation by
Pad\'e approximants works well in regions of the complex plane away from
branch-cut singularities, so we take all of the integration contours to be
straight lines at $\mathrm{arg}(\xi)=\pi/4$. The integrals computed in this
way are complex. The findings 
of the previous section suggest that by taking the sum as in
Eq.~\rf{eq.transs} one should be able to choose the imaginary part of the
constant $c$ so that the result is real for some range of $w$. 
%%%PRL (up to various errors discussed below). 
This is indeed the case and gives a value for $\mathrm{Im}(c)$
consistent with Eq.~\rf{csol} (with $p=-1$). A combined measure of error is
the imaginary part of the result of the resummation -- it remains very small
(below $0.01 \%$ relative to the real part) for $w>0.25$.

\begin{figure}[ht]
\includegraphics[height=0.32\textheight]{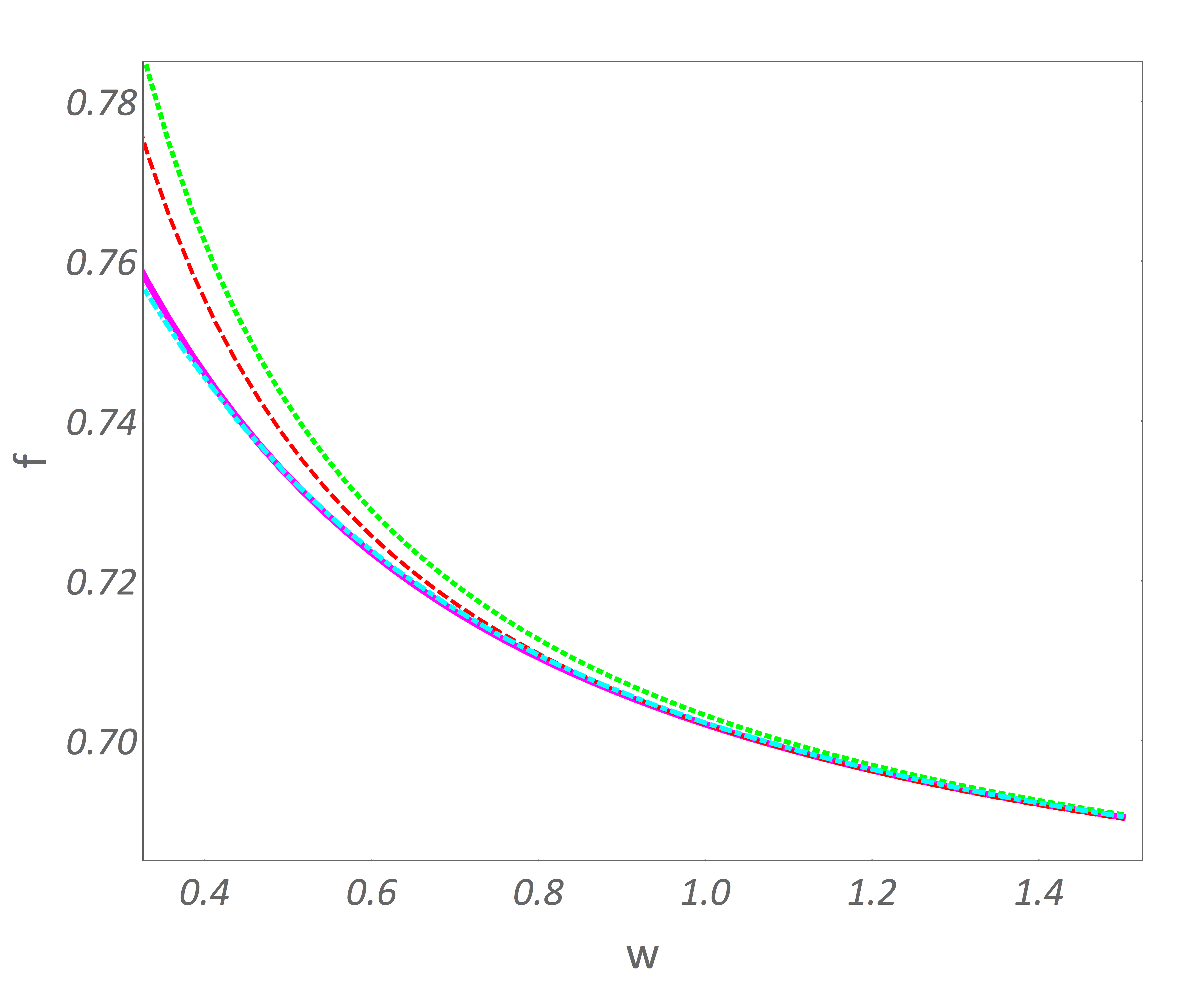}
\caption{The hydrodynamic attractor (magenta), compared with the
  resummation result (cyan, dot-dashed) and the gradient expansion of order
  1 (red, dashed) and 2 (green, dotted). } 
\label{fig.all}
\end{figure}

We compared the result of the resummation with the numerically computed
attractor, which required fitting the integration constant $r=0.049$ [see
Eq.~\rf{csol}]. 
As seen in Fig.~\ref{fig.all}, the generalized Borel sum of the
gradient series indeed follows the attractor.  
Note that to match the attractor we need to choose the 
coefficient $r\neq -\mathrm{Re}(c)$, 
while naively one might expect that the attractor should correspond to
omitting the exponential terms. 
This suggests that the resummation of the gradient 
series contains exponentially decaying terms which are not canceled by the
exponential terms from higher orders in the transseries. Consequently, 
instead of thinking about nonhydrodynamic modes in terms of perturbations
about the gradient expansion one should more properly think of them as
perturbations around the hydrodynamic attractor.

\vspace{10 pt}

\noindent \emph{Conclusions --} Recent advances in applying the theory of
resurgence to quantum
theories~\cite{Aniceto:2011nu,Argyres:2012vv,Argyres:2012ka,Dunne:2012ae,Aniceto:2013fka,Cherman:2014ofa}
have motivated us to 
rethink the foundational aspects of relativistic hydrodynamics. The root of
the problem is that causality precludes us from regarding hydrodynamics as a
truncated gradient expansion, yet the series itself is divergent. We propose 
to view hydrodynamics as an attractor which governs the late time behaviour of
systems in their approach to equilibrium. In the context of boost-invariant
flow in MIS theory we have constructed such an attractor in several ways which
are all 
consistent with each other.

By identifying the structure of singularities of the analytic continuation of
the Borel transform of the hydrodynamic series in terms of nonhydrodynamic
degrees of freedom, we hope that this Letter will provide a useful road map for
understanding the meaning of higher order gradients in the case of
\symm\ theory \cite{Heller:2013fn}. From the point of view of AdS/CFT, 
the exploratory studies described here suggest that the geometry constructed in
the gradient expansion of fluid-gravity duality should be viewed as the
leading term of a transseries containing the effects of quasinormal modes.

From a phenomenological perspective, MIS theory includes explicitly transport
coefficients for terms up to second order in gradients, but it generates a
gradient expansion to all orders.  The transport coefficients can only match
real QCD plasma up to second order. One may then wonder about the effects of
all of the higher order terms~\cite{Lublinsky:2007mm} which cannot be matched by
MIS theory. Our findings suggest that the attractor which
governs its late time behaviour is not very sensitive to the higher order
terms (even when the gradient series is resummed). This makes it less
surprising to learn that MIS theory can describe QGP evolution so well.

\vspace{10 pt}

% \noindent \emph{Acknowledgments --} 
We thank R.~Janik and
P.~Witaszczyk for their collaboration on related issues, A.~Cherman for very
valuable discussions and A.~Buchel, S.~Green, K.~Jensen, L.~Lehner,
M.~Lublinsky, 
M.~Martinez, W.~van~der~Schee for helpful feedback and correspondence. This
work was supported by the Polish National Science Centre grant
2012/07/B/ST2/03794. Research at Perimeter Institute is supported by the
Government of Canada through Industry Canada and by the Province of Ontario
through the Ministry of Research \& Innovation.

\newpage

\begin{appendix}

\section{Supplemental material: cancellation of ambiguities}

As stated in the section \emph{Resurgence} of the Letter, the key observation
is that the ambiguity at the leading order of the 
transseries is 
proportional to $\Omega$, so it can only be canceled by terms of order
$m=1$ or higher. The contributions to the transseries at orders $0-2$ can be
parameterized as 
\bea
f_R^{(0)} &=& \f{2}{3} +  \Omega b^{(0)}_1  + \Omega^2 b^{(0)}_2  + \dots,\nn\\
f_R^{(1)} &=& c\, \Omega\left(1 + \Omega b^{(1)}_1  + \dots\right), \nn\\
%f_R^{(2)} &=& c^2 \Omega^2 \left(-\f{3}{2} + \dots \right),
f_R^{(2)} &=& c^2 \Omega^2 \left(a_{2,0} + \dots \right),
\eea
where $b^{(m)}_n$ are constants, $a_{2,0} = -\frac{3}{2}+\frac{3
  C_{\lambda_{1}}}{2 C_{\eta}}$, and all the terms polynomial in $w$ have been 
dropped. The constants $b^{(m)}_n$ capture the leading contributions of the
ambiguous terms. Their values can be obtained as described in the 
section  \emph{Hydrodynamic Gradient Expansion at High Orders} in the
Letter. The result is 
\beal{eq.b}
b^{(0)}_1 &=& e^{i \pi \gamma p} \Gamma(\gamma+1)
h^{(0)}_1(\xi_{0}),\nn\\
b^{(0)}_2 &=& e^{2 i \pi \gamma p} \Gamma(2 \gamma+1)
h^{(0)}_2(2 \xi_{0}),\nn\\
b^{(1)}_1 &=& e^{i \pi \gamma p} \Gamma(\gamma+1)
h^{(1)}_1(\xi_{0}).
\eea
The odd integer $p$ appearing in the phase factor could a priori be different in
each function, but for a cancellation of the ambiguity to be possible they
need to be equal as written above. Note also that the superscript indicates 
the order in the transseries, so for example $h^{(0)}_k$ is what in the
section  \emph{Hydrodynamic Gradient Expansion at High Orders} section was denoted by $h_k$.
As stated in the Letter, the cancellation fixes the constant~$c$ up to a real
number shift. Alternatively, one may say that the cancellation of ambiguities
determines the imaginary part of~$c$.

Once the imaginary part of $c$ is determined, conditions at higher orders in
$\Omega$ should be satisfied automatically. At second order one finds
\beal{consicon}
2 \,a_{2,0} \,h^{(0)}_1(\xi_{0})- h^{(1)}_1(\xi_{0}) = 0,\nn\\
\Gamma(2\gamma+1) h^{(0)}_2(2 \xi_{0}) - a_{2,0} \Gamma(\gamma+1)^2
h^{(0)}_1(\xi_{0})^2 = 0.
\eea
Checking this explicitly is not easy, as it requires a very accurate numerical
calculation of 
the numbers $h^{(0)}_1(\xi_{0}), h^{(1)}_1(\xi_{0})$ and
$h^{(0)}_2(2\xi_{0})$. We found\footnote{For technical reasons, we set
  $C_{\lambda_{1}} = 0$ while performing the check.} that the first of the
consistency conditions Eq.~\rf{consicon} is 
satisfied at the 
level of $5\%$. The second condition in Eq.~\rf{consicon} is much harder to
check given the level of accuracy attainable using Pad\'e
approximants in the vicinity of branch cut singularities. 

Note finally that the term
involving $h^{(0)}_2$ would be absent if 
we did not include the singularity at $\xi=2\xi_0$ in Eq.~(19) in the Letter, but its
presence is necessary for the cancellation mechanism outlined here. We expect
a similar mechanism to operate at higher orders. These expectations are also
supported by the success of the resummation described in section \emph{Resummation} in the Letter. It is
also reassuring that the value of $c$ determined there matches the value given in Eq.~(23) in the Letter.

\end{appendix}

\bibliography{mis}{}
\bibliographystyle{utphys}

\end{document}